# The origin of bias independent conductance plateaus and zero bias conductance peaks in $Bi_2Se_3$/$NbSe_2$ hybrid structures

Hui Li[1]*, Tong Zhou[1]*, Jun He[1], Huanwen Wang[1], Huachen Zhang[1], Hong-Chao Liu[1], Ya Yi[1], Changming Wu[1], Kam Tuen Law[1]**, Hongtao He[2]**, Jiannong Wang[1]**

[1]*Department of Physics, the Hong Kong University of Science and Technology, Clear Water Bay, Hong Kong, China*

[2]*Department of Physics, South University of Science and Technology of China, Shenzhen, Guangdong 518055, China*

* These authors contributed equally to this work.

** Correspondence and requests for materials should be addressed to K.T. L. (email: phlaw@ust.hk), H. H. (email: heht@sustc.edu.cn) and J. W. (email: phjwang@ust.hk).

## Abstract

Superconducting proximity effect (SPE) in topological insulator (TI) and superconductor (SC) hybrid structure has attracted intense attention in recent years in an effort to search for mysterious Majorana fermions (MFs) in condensed matter systems. Here we report on the SPE in a $Bi_2Se_3$/$NbSe_2$ junction fabricated with an all-dry transfer method. Resulting from the highly transparent interface, two sharp resistance drops are observed at 7 K and 2 K, respectively, corresponding to the superconducting transition of $NbSe_2$ flake and the SPE induced superconductivity in $Bi_2Se_3$ flake. Experimentally measured differential conductance spectra exhibit a bias-independent conductance plateau (BICP) in the vicinity of zero bias below 7 K. As temperatures further decrease a zero bias conductance peak (ZBCP) emerges from the plateau and becomes more enhanced and sharpened at lower temperatures. Our numerically simulated differential conductance spectra reproduce the observed BICP and ZBCP and show that the SPE in topological surface states (TSS) is much stronger than that in the bulk states of $Bi_2Se_3$. The SPE induced superconducting gap for the TSS of $Bi_2Se_3$ is comparable to that of $NbSe_2$ and gives rise to the observed BICP below 7 K. In contrast, the SPE induced superconducting gap for the bulk states of Bi$_2$Se$_3$ is an order of magnitude smaller than that of NbSe$_2$ and superconducting TSS. These weakly paired bulk states in Bi$_2$Se$_3$ give rise to the ZBCP below 2 K. Our study has clearly unveiled the different roles of TSS and bulk stats in SPE, clarified the physical origin of the SPE induced features, and shined light on further investigation of SPE and MF in TI/SC hybrid structures.

Majorana fermions are long-sought fundamental particles, whose antiparticles are identical to themselves.[1] In condensed matter systems, Majorana fermions follow non-abelian statistics, and have great potential applications in fault-tolerant topological quantum computation.[2,3] Motivated by the promising applications, much effort has been devoted to the search for the signature of Majorana fermions.[4-9] Recently, theoretical work has predicted that the Majorana fermions might reside in the interface of the topological insulators and *s*-wave superconductors (TIs/SC) hybrid structures.[10-14] According to theoretical predictions, when TIs contact with *s*-wave superconductors, the topological superconductor (TSC) with effectively spinless $p_x+ip_y$ pairing would be formed through the superconducting proximity effect (SPE), and Majorana fermions are expected to emerge in Abrikosov vortex cores.

The search for Majorana fermions in TIs has stimulated an extensive transport study of SPE in various SC/TI/SC and SC/TI junctions. The existence of supercurrent,[15] multiple Andreev reflection,[15] and ac/dc Josephson effect[16,17] have been revealed in SC/TI/SC junctions. As for SC/TI junctions with high barriers at the interface, electron tunneling would suppress the differential conductance across the junction within the superconducting gap, giving rise to a V-shaped differential conductance spectrum.[18] But for SC/TI junctions with transparent interfaces, Andreev reflections would enhance the differential conductance within the superconducting gap, leading to a differential conductance plateau which is almost twice as high as the differential conductance in the normal state.[19] Another intriguing phenomenon observed in SC/TI junctions is the presence of a zero bias conductance peak (ZBCP) in the differential conductance spectra, since it might be a transport signature of topological superconductors.[20] The ZBCP observed in Cu doped $Bi_2Se_3$[21] and in In doped SnTe[22] have been associated with topological superconductivity.[23]

In this Letter, we report the magnetotransport studies of a $Bi_2Se_3$/$NbSe_2$ hybrid structure with a transparent interface. Two resistance drops at about 7 K and 2 K are observed, which originates from the superconducting transition of $NbSe_2$ flakes and the SPE induced superconductivity in $Bi_2Se_3$ flakes, respectively. In differential conductance spectra a bias-independent conductance plateau (BICP) in the vicinity of zero bias is observed below ~7 K while a sharp ZBCP is emerged below ~2 K due to the transparent interface between the $Bi_2Se_3$ and $NbSe_2$ flakes. Our numerical simulation of the differential conductance spectra of our hybrid structure can reproduce the experimentally observed BICP and ZBCP. Detailed analyses indicate that the occurrence of the BICP is a result of the Andreev reflections mediated by the SPE induced superconducting TSS of $Bi_2Se_3$, while the sharp ZBCP arises from the SPE induced superconducting bulk states of $Bi_2Se_3$. Moreover, we found that the SPE induced superconducting gap of TSS is one order of magnitude larger than that of bulk states and comparable to the superconducting gap of $NbSe_2$, indicating the SPE of TSS is much stronger than that of bulk states in our $Bi_2Se_3$/$NbSe_2$ structure. Our experimental and theoretical results unveil the origin of the BICP[17] and ZBCP in SC/TI hybrid structures and provide a clear understanding in the proximity effects of superconductors on TIs.

$Bi_2Se_3$/$NbSe_2$ hybrid structures were fabricated based on an all-dry transfer method.[24] The $NbSe_2$ flakes and $Bi_2Se_3$ flakes were mechanically exfoliated from the high-quality single crystals, and were transferred to Si substrates with a 300-nm-thick $SiO_2$ and PDMS-based gel supplier, respectively. Using micromanipulator and digital camera, the $Bi_2Se_3$ flakes were further transferred onto the selected $NbSe_2$ flakes for device fabrication using e-beam lithography. Au/Cr electrodes with the thickness of 275 nm/25 nm were deposited via thermal evaporation. The optical image of a fabricated device was illustrated in Fig. 1(a). A four terminal measurement across the $Bi_2Se_3$/$NbSe_2$ junction was performed in a PPMS Dilution Refrigerator system with the temperature down to 100 mK. The Resistance versus Temperature ($R$-$T$) curves were measured using Keithley 6221 and 2182A as the current source and voltmeter, respectively. Keithley 6221 and Lock-in amplifiers with a frequency of 741 Hz was adopted for the differential conductance, $dI/dV$, spectra measurements.

Prior to putting the devices into the PPMS Dilution Refrigerator, we first measured the $R$-$T$ curves with $T > 2$ K in PPMS base chamber at different magnetic fields indicated, as shown in Fig. 1(b). The resistance of the $Bi_2Se_3$/$NbSe_2$ junction first increases as temperature decreases, and suddenly drops at a critical temperature $T_c \sim 7$ K, corresponding to the superconducting phase transition of the $NbSe_2$ flake. As the applied magnetic field increases, the $T_c$ decreases from ~ 7 K at zero magnetic field to temperatures lower than 2 K at 4 T. Interestingly, besides this sharp drop of the resistance at $T_c$ ~7 K, a second resistance drop starts to appear below 3 K at zero magnetic field, which is attributed to the SPE induced superconductivity in the topological insulator $Bi_2Se_3$ flake. The observation of the SPE induced superconductivity in the hybrid structure suggests that the interface is of high transparency in our devices.[19]

In order to investigate in details the SPE induced superconductivity in $Bi_2Se_3$ flake, we focus on the magneto-transport properties and differential conductance spectra of the $Bi_2Se_3$/$NbSe_2$ junction at low temperatures ranging from 0.1 K to 4 K. Fig. 1(c) shows the measured *R-T* curves for 0.1 K < *T* < 4 K at different magnetic fields with resistances normalized to the values at *T* = 3.5 K. At zero magnetic field, with decreasing temperature the resistance decreases slowly first and then starts to drop quickly below ~ 2 K. $T_{in}$ ~ 2 K corresponds to the transition temperature of the SPE induced superconductivity in $Bi_2Se_3$ flake. The ratio of *R*(0.1 K)/*R*(3.5 K) is about 0.84 at zero magnetic field. By applying magnetic fields this ratio increases to 0.97 at 1.5 T, and $T_{in}$ shifts to lower temperatures as expected.

Magnetoresistance (MR) of the junction at different temperatures as indicated are shown in Fig. 1(d). Two important MR features are noticed at the lowest *T* of 0.1 K, one forming a sharp MR dip at low magnetic field region and the other forming a broader resistance valley at high magnetic field region, revealing the two critical magnetic fields for the superconducting phases existed in this hybrid structure. As temperature is increased, the sharp MR dip is quickly suppressed at ~ 2 K, which coincides with the critical temperature $T_{in}$ of the $Bi_2Se_3$ flake observed in *R-T* curves shown in Fig. 1(c). At the same time, the broad resistance valley shrinks significantly from ~ 5 T at 0.1 K to 2 T at 4 K, which are consistent with the superconductivity transition of $NbSe_2$ flake observed in *R-T* curves shown in Fig. 1(b). These results indicate the suppression of the Cooper pair formation and thus the destruction of the superconductivity at high temperatures and high magnetic fields.

The above *R-T* and MR results clearly demonstrate the occurrence of the SPE in our $Bi_2Se_3$/$NbSe_2$ junction. To gain more insight into the effect, we have performed the four-terminal differential conductance spectroscopy measurements of the junction. Fig. 2(a) shows the measured differential conductance spectra (*dI/dV vs. V*) at temperatures from 0.5 K to 4 K. One can see that a BICP appears in the vicinity of zero bias. As temperatures decrease below 2 K, a very sharp ZBCP is emerged and superimposed on the BICP. We will focus on the behavior of the BICP first. As shown in Fig. 2(a), the height of the flat conductance plateau changes little from 0.5 K to 4 K, but its width decreases obviously with increasing temperature. In Fig. 2(b) the width of the flat conductance plateau against measurement temperature is plotted as black solid dots and the temperature dependence of $NbSe_2$ superconducting gap based on BCS theory is plotted as solid red line for reference. Its width at 0.5 K is estimated to be ~ 2 meV, which is comparable to the superconducting gap of NbSe$_2$ flake estimated from the BCS theory 2Δ(0) = 3.52$k_B T_c$ ~ 2.12 meV. As temperature increases the width is reduced due to suppressed superconductivity of NbSe$_2$ flake. We note that the width reduction is faster than that expected from the BCS theory. This deviation may be caused by the anisotropy and Fermi-surface-sheet dependence of the superconducting gap of NbSe$_2$.[25,26] Besides the strong temperature dependence, the width of the plateau also shows strong dependence on the applied magnetic field. Fig. 2(c) shows the $dI/dV$ spectra at different magnetic fields with a constant temperature of 0.5 K. It is clear that the plateau width is greatly reduced by increasing magnetic field (see Fig. 2(d)) and completely suppressed at ~ 5 T, in agreement with the critical magnetic field of NbSe$_2$ flake measured in Fig. 1(d).

Generally speaking, Andreev reflection is expected to greatly enhance the conductance across a normal metal and superconductor junction with an electrically transparent interface. A BICP within the superconducting gap Δ would be thus observed around zero voltage bias in the differential conductance spectra.[27-29] The evident BICP shown in Fig. 2(a) indicates that an electrical transparent interface has been achieved in our Bi$_2$Se$_3$/NbSe$_2$ device. However, since the Fermi level of our Bi$_2$Se$_3$ flake lies in the conduction band (See supplementary S1), either TSS or bulk electrons can be responsible for the BICP. As the wave function of TSS electrons is more localized at the interface than that of bulk electrons, it is very likely that superconducting transition by the SPE occurs first in TSS once NbSe$_2$ flake becomes superconducting. Indeed, in our detailed numerical analysis presented later, we show that the BICP is solely due to the superconducting TSS and the proximity gap induced on the TSS by the NbSe$_2$ is almost as big as the superconducting gap in NbSe$_2$.

We now turn the attention to the ZBCP emerged below 2 K. Fig. 3 (a) shows the measured differential conductance spectra at different temperatures within a smaller bias range of ±1 mV. The ZBCP is pronounced at 1.5 K and becomes stronger and sharper as temperature decreases. The emergence of the ZBCP coincides with the second resistance drop observed in $R$-$T$ curves (see Fig. 1(b) and 1(c)). This means the ZBCP is associated with the SPE induced superconductivity in Bi$_2$Se$_3$ flake. The temperature dependence of the ZBCP intensity and width (i.e. full-width at the half-maximum) is plotted in Fig. 3(b). As it can be seen, the ZBCP intensity is first reduced slowly as temperature increases from 0.1 K to 0.5 K, then faster as temperature further increases, and finally is immeasurable around 2 K. More interestingly, the ZBCP width is almost constant below 0.5 K, which has a value of about 0.13 meV about 10% of 2Δ(0), and then is quickly broadened above 1 K. Such behaviors are in sharp contrast to the case of the BICP (see Fig. 2(b)), where the height changes little but the width decreases obviously with increasing temperatures. This indicates that the ZBCP width is determined by thermal energy and the energy magnitude involved in the ZBCP is at the same order as the thermal energy. For any realistic system, the finite temperature would make the Andreev reflections decoherent over a short length, and thus the short lifetime. Considering the Heisenberg's uncertainty principle, $\Delta E \Delta \tau \sim \hbar$, the effective energy broadening term would be induced leading to the overall broadening of the conductance spectrum for the ZBCP.[30, 31] The magnetic field dependence of the ZBCP at 0.1 K is illustrated in Fig. 3(c). With increasing applied magnetic field the width of the ZBCP is broadened and the intensity is reduced. Fig. 3(d) plots the magnetic field dependence of the ZBCP intensity and width. The ZBCP intensity shows a very strong dependence on the magnetic field and decreases rapidly with the applied magnetic field. When the magnetic field is above 1 T, the ZBCP cannot be measured, indicating the proximity induced superconductivity in $Bi_2Se_3$ flakes is quenched, in agreement with the critical magnetic field of $Bi_2Se_3$ flake measured in Fig. 1(d).

The ZBCP has been reported in various S-N hybrid structures,[15,20,32-36] but its physical origin is still under debate. The ZBCP can be induced by the pair current across superconductor-semiconductor interfaces.[37] But the ZBCP width is expected to increase with decreasing temperatures, which is contrary to the thermal broadening behavior observed in our experiment. The ZBCP can also arise from the incoherent accumulation of Andreev reflection and it begins to appear in the differential conductance spectra right below $T_c$.[38] However, the ZBCP in our devices can only be observed below 2 K, which is much lower than $T_c$. Furthermore, because of the phase conjugation between electrons and holes, coherent quantum interference has been proposed to account for the ZBCP in SC/SM junctions with low transparent interface.[39] However, for our system, the interface is highly transparent and the strong spin orbital coupling in $Bi_2Se_3$ flakes further prevents the electron-hole phase conjugation. Finally, the Andreev bound state of an anisotropy superconductor[28,39-41] or the Majorana zero mode in the core of the vortices of topological superconductor[12,42] can also give rise to the ZBCP. ZBCPs of these kinds usually show a weak dependence on the magnetic field. However, these are apparently not our case. The ZBCP observed is quenched quickly by small external magnetic field (see Fig. 3(c)).

To summarize the experimentally measured differential conductance spectra at low temperature and zero magnetic field exhibit two distinct features: one is the sharp ZBCP with its width approximately 10% of 2Δ(0) and the other is the BICP with its energy range comparable to 2Δ(0). In order to understand the physical origins of both the ZBCP and the BICP, we theoretically calculate the *dI/dV* spectra for our $Bi_2Se_3$/$NbSe_2$ junction. In our theoretical calculations (see Supplementary S2), we simulate the TI-SC junction with the schematic set-up in Fig. 4(a), which realistically represents our device structure, where the chemical potential of the TI is just touching the bulk conduction band bottom (e.g., $E_F = \mu_1$, indicated by the red dashed line in the inset of Fig.4 (b)). In this case, the Fermi surface consists of multiple Fermi circles, where the outermost Fermi circle derived from the TSS encloses the inner one derived from the bulk states. Therefore, both the bulk states and the TSS from $Bi_2Se_3$ would participate in the SPEs and drive currents across the TI-SC junction via Andreev reflections. The *dI/dV* spectra from our numerical simulations with $E_F = \mu_1$ is shown in Fig. 4(b). Remarkably, both features of BICP and ZBCP are reproduced in a qualitatively excellent agreement with the experimental data (see Fig. 2(a) and Fig. 3(a)). More importantly, by increasing the number of quintuple layers *N* of $Bi_2Se_3$, the sharp ZBCP is enhanced, while the BICP remains almost unaffected. Their contrasting dependence on *N* suggests that the ZBCP may be associated with the bulk states near the conduction band minimum, while the BICP likely originates from the TSS.

To further verify the physical origins of the peak and the plateau, we calculate the *dI/dV* spectra for our $Bi_2Se_3$/$NbSe_2$ junction in another two separate cases. First, we study the case where only the TSS in $Bi_2Se_3$ is involved in the SPE. By setting the chemical potential in $Bi_2Se_3$ to lie within the bulk gap ($E_F = \mu_2$ indicated by the green dashed line in the inset of Fig. 4(c)), we exclude the contributions from the bulk states and the Andreev reflections at low-voltage bias are driven by the TSS only. As shown in Fig. 4(c), the ZBCP vanishes, while the BICP remains intact and its height is almost unchanged with the quintuple-layer number *N*. Therefore, we confirm that the BICP indeed originates from the SPE in the TSS. Second, we study the case where only the bulk states near the conduction band bottom of $Bi_2Se_3$ are filled at the Fermi energy. This is done by tuning the tight-binding parameters of $Bi_2Se_3$ such that it becomes a trivial band insulator with $E_F = \mu_3$ as shown in the inset of Fig. 4(d) indicated by the blue dashed line. In this case, the BICP disappears in the $dI/dV$ spectra with only a sharp ZBCP left (Fig. 4(d)), confirming that the sharp ZBCP arises from the SPE in the bulk states near the conduction band bottom.

Their contrasting signatures in the $dI/dV$ spectra suggest that the bulk states and TSS in $Bi_2Se_3$ behave very differently in the SPEs. To understand their distinct roles in the SPEs, we calculate the spectral function of $Bi_2Se_3$ with $E_f = \mu_1$ placed right on top of superconducting $NbSe_2$, as shown in Fig. 4 (e). Evidently, the outermost Fermi circle associated with the TSS acquires a sizable induced gap $\Delta'_S$ ($2\Delta'_S$ is indicated by the double arrow in yellow in Fig. 4(e)) that is comparable to the parent superconducting gap $\Delta$. This result is consistent with experimentally obtained width of the BICP at low temperature (see Fig. 2(b)). In contrast, the induced gaps on the inner Fermi circles $\Delta'_B$ ($2\Delta'_B$ is indicated by the double arrow in white in Fig. 4(e)) are an order of magnitude smaller in size, which is also in agreement with experimentally obtained ZBCP width (see Fig. 3(b)). In other words, the SPE in the TSS are much stronger comparing to the SPEs in the bulk states. The strong SPE and large induced gap in the TSS give rise to the BICP observed in our experiment at temperatures a bit below the host $T_c$ (see Fig. 2(b) dashed curves, estimated as 6.5 K). On the other hand, the weak SPE and small induced gap in the bulk states are responsible for the ZBCP observed below 2 K, which is much lower than the host $T_c$.

In addition, we note that some differential conductance ripples appear outside the BICP in Fig. 2(a) and the ZBCP in Fig. 3(a). Similar ripples were observed in differential conductance spectra outside the superconducting gap region in the study of an $Al-Al_2O_3-Pb$ tunneling junction.[43] The phenomenon was ascribed to the multi-phonon effect, revealing the important role of phonon interactions in the superconductivity of Pb. Besides, in the study of $UPd_2Al_3-AlO_x-Pb$ tunneling junctions, the observation of such ripples was associated with some magnetic excitations, which is essential to the emergence of unconventional superconductivity in heavy fermion compound $UPd_2Al_3$.[44] At present stage, the physical origin of the ripples observed in our work is still unknown and needs further exploration. But these ripples show the same temperature and magnetic field dependence as the BICP or the ZBCP, indicating the close relationship between the ripples and the SPE induced superconductivity in TSS or bulk states of $Bi_2Se_3$.

In conclusion, the SPE is investigated in $Bi_2Se_3$/$NbSe_2$ hybrid structure with transparent interface. Two distinct features in the measured differential conductance spectra are revealed: one is the BICP observed once $NbSe_2$ becomes superconducting and the other is the sharp ZBCP emerged at much lower temperature. Both features are reproduced in our numerical simulations. We identify that the ZBCP originates from the SPE in the bulk states, while the pronounced BICP results from the SPE in the TSS. The *dI/dV* spectra from our transport measurements provide the strong evidence for a prominent SPE on the TSS in our $Bi_2Se_3$/$NbSe_2$ junction. This suggests our $Bi_2Se_3$/$NbSe_2$ junction can be a promising candidate for realizing Majorana fermions at the vortex core on the surface of a 3D topological insulator. Our work not only provides new insight into the different roles of TSS and bulk states in the SPE of TIs, but also proposes a new physical mechanism for ZBCP, which would have some implications to the search for Majorana fermions in TI and superconductor hybrid structures.

## Acknowledgements

This work was supported in part by the Research Grants Council of the Hong Kong SAR under Grant Nos. 16305215, AOE/P-04/08-3-II, and HKU9/CRF/13G-1, and in part by the National Natural Science Foundation of China under Grant No. 11574129, and in part by the Special Funds for the Development of Strategic Emerging Industries in Shenzhen, Grant No. JCYJ20140714151402765. The electron-beam lithography facility is supported by the Raith-HKUST Nanotechnology Laboratory at MCPF (SEG HKUST08).

**Figure captions:**

FIG. 1. (Color online) (a) Optical image of the $Bi_2Se_3/NbSe_2$ hybrid structure devices. (b) Normalized *R-T* curves of the $Bi_2Se_3/NbSe_2$ hybrid structures at different applied magnetic fields above 2 K and (c) between 100 mK and 4 K, showing the superconducting phase transition of $NbSe_2$ and SPE-induced superconductivity in $Bi_2Se_3$. (d) Normalized *R-B* curves of the $Bi_2Se_3/NbSe_2$ hybrid structures measured at various temperatures.

FIG. 2. (Color online) (a) Differential conductance spectra *dI*/*dV* of the $Bi_2Se_3/NbSe_2$ hybrid structures at temperatures ranging from 0.5 K to 4 K and zero magnetic field. (b) The BICP width as a function of temperatures as indicated by black solid dots. For comparison, the temperature dependent superconducting gap of $NbSe_2$ calculated by the BCS theory with three different $T_c$ is also plotted. (c) Differential conductance spectra *dI*/*dV* at different magnetic fields and 0.5 K. (d) The BICP width as a function of magnetic field at 0.5 K is shown as black solid dots and the solid line is a guide to the eyes.

FIG. 3. (Color online) (a) Differential conductance spectra *dI*/*dV* of the $Bi_2Se_3/NbSe_2$ hybrid structures at temperatures ranging from 0.1 K to 3 K and zero magnetic field. (b) The ZBCP peak intensity (black open circles) and width (blue open squares) as a function of temperatures at zero magnetic field. Solid lines are guides to the eyes. (c) Differential conductance spectra *dI*/*dV* at different magnetic fields and 0.1 K. (d) The peak intensity (black open circles) and width (blue open squares) as a function of magnetic field at 0.1 K. Solid lines are guides to the eyes.

**FIG. 4**. (Color online) Numerical simulations of the *dI/dV* spectra and the superconducting proximity effects for the $Bi_2Se_3$/$NbSe_2$ junction. The insets in (b)-(d) show the energy spectrum of $Bi_2Se_3$ near the $\Gamma$-point in the surface Brillouin zone. The dashed-lines indicate the locations of the Fermi energy in each individual case. (a) Schematic showing the Andreev reflections and the SPEs in the TI-SC junction. (b)-(c) Zero-temperature *dI/dV* spectra for topologically nontrivial $Bi_2Se_3$ with different numbers of quintuple layers N=9, 12, 15, 20. The Fermi energy of $Bi_2Se_3$ in (b) is set to be $E_F = \mu_1$ (see inset) such that both the surface states and the bulk states near the conduction band minimum are filled. Both features of the ZBCP and BICP are reproduced in qualitatively good agreement with the experimental data. The Fermi energy of $Bi_2Se_3$ in (c) is set to be $E_F = \mu_2$ (see inset) such that only the surface states are filled. The ZBCP disappears while the flat plateau remains almost unaffected by the number of quintuple layers N. (d) Zero-temperature *dI/dV* spectra for topologically trivial $Bi_2Se_3$ with N=20. The Fermi energy of $Bi_2Se_3$ is set to be $E_F = \mu_3$ (see inset) such that only the bulk conduction band minimum is filled. The ZBCP remains while the BICP disappears. (e) The spectral weight of the bottom quintuple layer of a section of TI placed right on top of superconducting $NbSe_2$ thin flakes. The Fermi energy of $Bi_2Se_3$ is the same as in (b). The double-arrows in yellow/white indicate twice of the induced gaps $2\Delta'_S/2\Delta'_B$ on the surface/bulk states.

Evidently, $\Delta'_S$ is significantly greater than $\Delta'_B$.

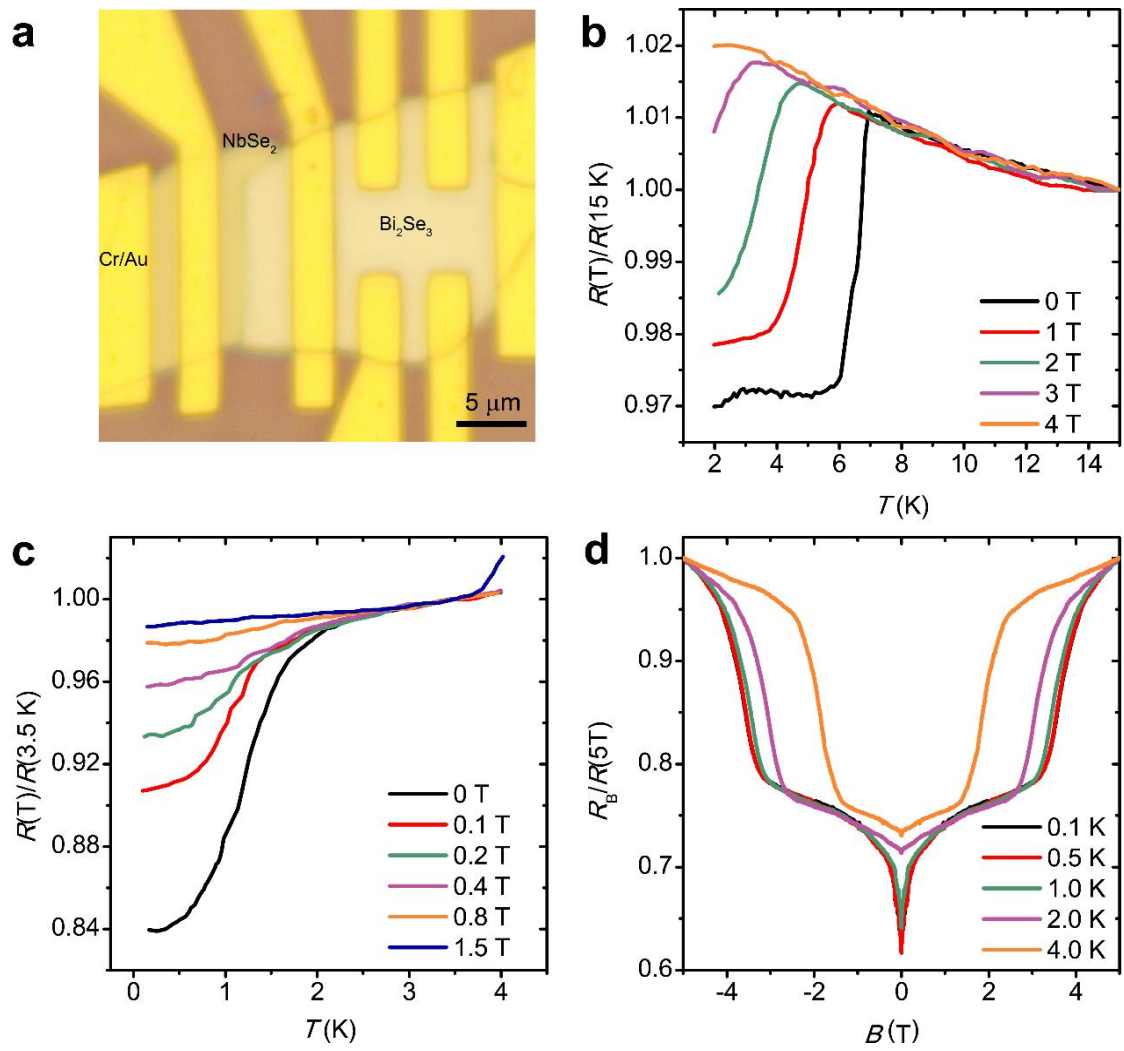

**Fig. 1**

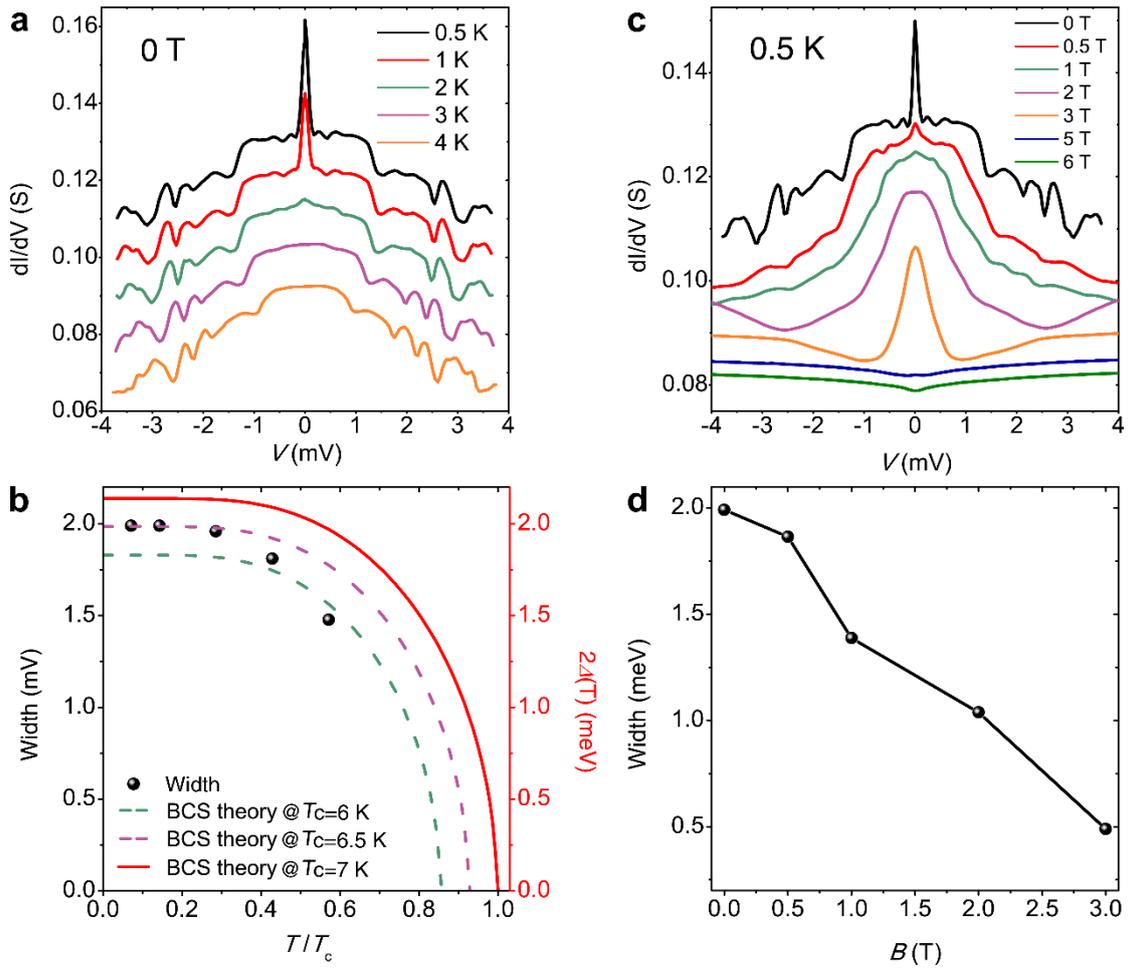

**Fig. 2**

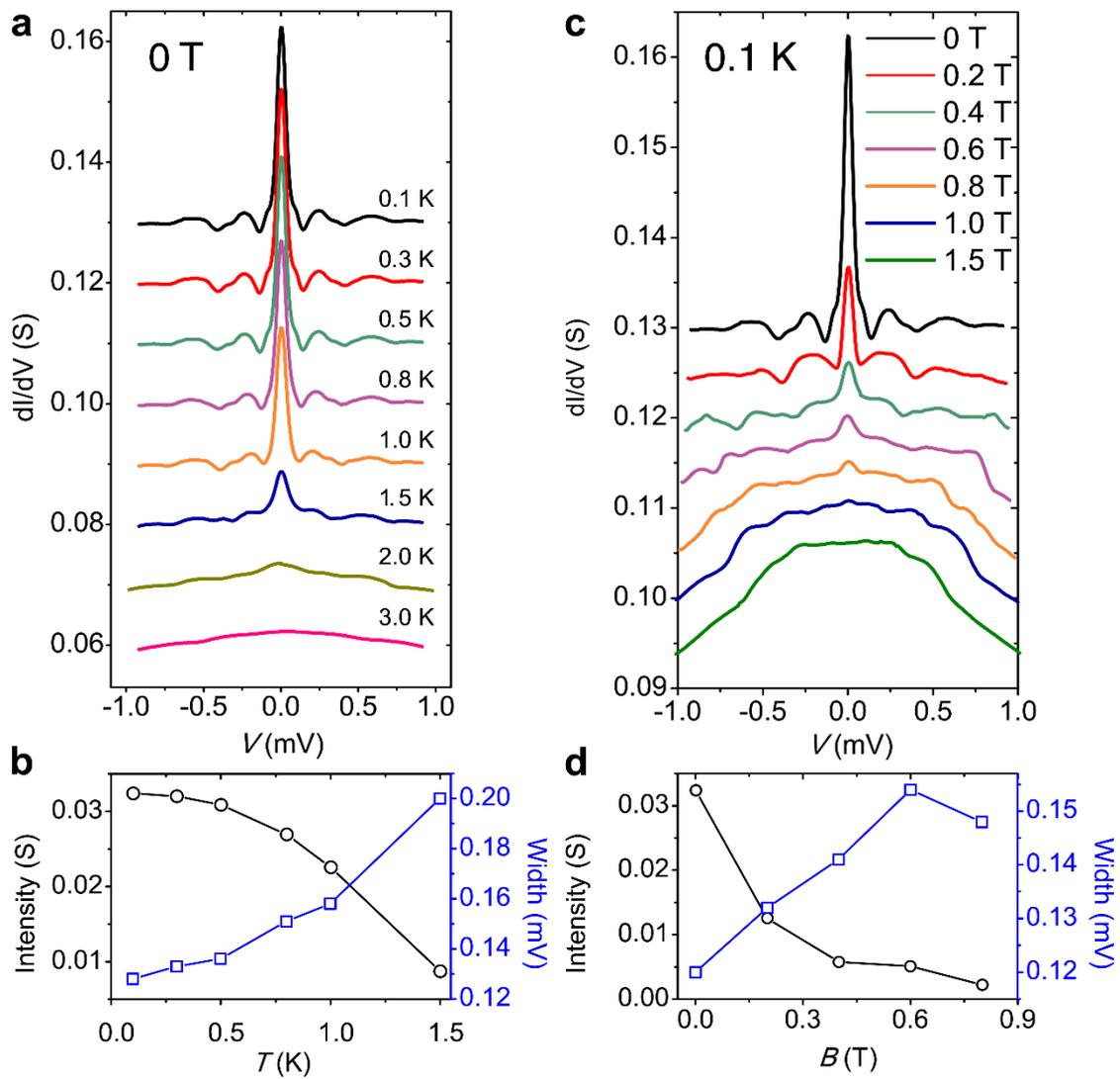

**Fig. 3**

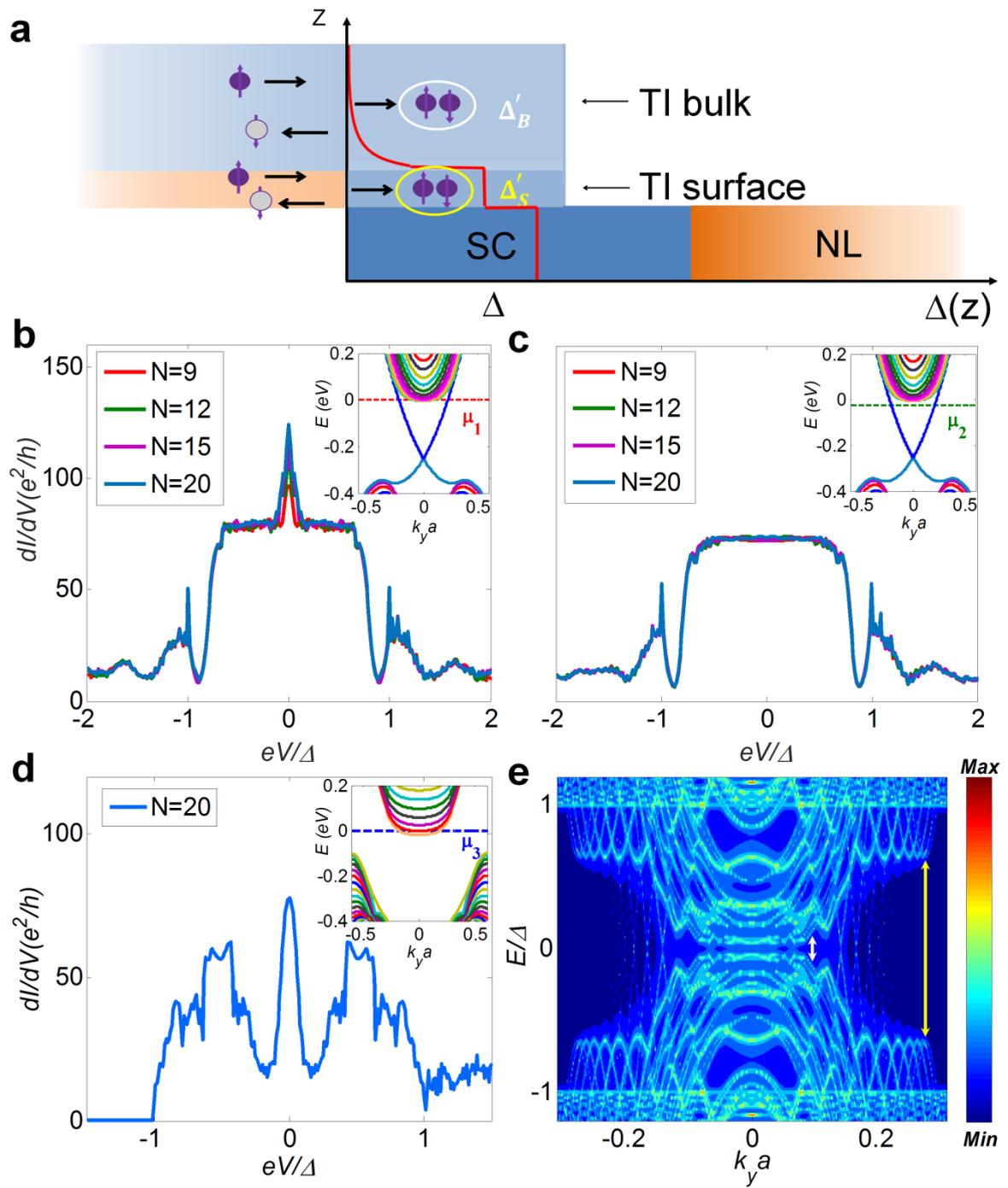

**Fig. 4**